\documentclass[aps,prb,superscriptaddress,showpacs,twocolumn,floatfix]{revtex4-1}

\usepackage{graphicx}
\usepackage{bm}
\usepackage{amssymb}

\usepackage{epsfig}

\usepackage{amsmath}

\usepackage{pstricks}

\def\va1{\vec{a}_{1}}

\def\vb1{\vec{b}_{1}}

\def\vd1{\vec{\delta}_{1}}

\newcommand{\ba}{\begin{eqnarray}}
\newcommand{\ea}{\end{eqnarray}}
\def\be{\begin{equation}}
\def\ee{\end{equation}}

\def\vk{\vec{k}}
\def\vx{\vec{x}}
\def\vy{\vec{y}}

\def\vra{\vec{r}_{\alpha}}
\def\vrb{\vec{r}_{\beta}}

\def\vk{\mathbf{k}}

\newcommand{\op}[1]{ \hat{#1}}
\newcommand{\ve}[1]{ \mathbf{#1}}
\newcommand{\bimn}{Bi$_3$Mn$_4$O$_{12}$(NO$_3$)}

\begin{document}


\title{Exotic disordered phases in the quantum $J_1-J_2$ model on the honeycomb lattice}

\author{Hao Zhang}
\email{Corresponding author: zhanghao@issp.u-tokyo.ac.jp}
\affiliation{Institute for Solid State Physics, University of Tokyo,
Kashiwa, Chiba 277-8581, Japan}
\author{C.\ A.\ Lamas}
\affiliation{Laboratoire de Physique Th\'eorique, IRSAMC, CNRS and Universit\'e de Toulouse, UPS,
 F-31062 Toulouse, France}

\begin{abstract}

We study the ground-state phase diagram of the frustrated quantum
$J_1-J_2$ Heisenberg antiferromagnet on the honeycomb lattice using
a mean field approach in terms of the Schwinger boson representation
of the spin operators. We present results for the ground-state
energy, local magnetization, energy gap and spin-spin correlations.
The system shows magnetic long range order for $0\leq
J_{2}/J_{1}\lesssim 0.2075$ (N\'eel) and $0.398\lesssim
J_{2}/J_{1}\leq 0.5$ (spiral). In the intermediate region, we find
two magnetically disordered phases: a gapped spin liquid phase which
shows short-range N\'eel correlations $(0.2075 \lesssim J_{2}/J_{1}
\lesssim 0.3732)$, and a lattice nematic phase $(0.3732 \lesssim
J_{2}/J_{1}\lesssim 0.398)$, which is magnetically disordered but
breaks lattice rotational symmetry. The errors in the values of the
phase boundaries which are implicit in the number of significant
figures quoted, correspond purely to the error in the extrapolation
of our finite-size results to the thermodynamic limit.

\end{abstract}

\maketitle

\section{Introduction}

The two-dimensional (2D) Heisenberg model on bipartite lattices has
been intensively studied in the last years. In the unfrustrated
case, the classical ground state is obtained when all the spins in
one sublattice are  pointing in a given direction whereas in the
other sublattice the spins are pointing in the opposite direction.
However, in the quantum case this state is not the real ground
state, in fact this is not an eigenstate of the Hamiltonian. The
quantum ground state is exactly known in one dimension\cite{Bethe},
but no exact results for the two dimensional antiferromagnet are
known, even for simple lattices as the square lattice. However,
several experimental and numerical studies suggested that the ground
state is in fact the spin SU(2) symmetry broken N\'eel type state.
In contrast, when we include frustration in the system, for example
by including second neighbor interactions, the ground state may
become much more complicated.

In the quantum case, the ground state energy is lower than the classical value, due to the quantum
fluctuations. The effects of these fluctuations vary depending on the dimension, the spin
 quantum number,  the presence of frustrating interactions and the coordination number of the lattice.
One can ask what the quantum fluctuations are when the coordination
number is changed. In two dimensions two paradigmatic examples of
unfrustrated systems are the square lattice, with coordination
number
 $z=4$, and the honeycomb lattice with $z=3$. Previous results\cite{Hamer,Sanvik_1997} have shown that
the staggered magnetization is smaller in the $z=3$ case. This
behavior is in accord with the tendency
 towards a less classical behavior for systems of lower coordination number.

The inclusion of frustration in 2D quantum antiferromagnets is expected to enhance the
 effect of quantum spin fluctuations and hence
suppress magnetic order \cite{Anderson}. This idea has motivated
 many researchers to look for its realization \cite{Sachdev1,Sachdev2,Sandvik,Moessner,Poilblanc}.
A special scenario to check this is the frustrated Heisenberg model
on the honeycomb lattice. Due to the small coordination number
$(z=3)$ which is the lowest allowed in a 2D system, quantum
fluctuations could be expected to be stronger than those on the
square lattice and may destroy the antiferromagnetic
order\cite{Mattsson,Takano,Hermele,Kumar}.

\begin{figure}[t!]
\includegraphics[width=0.35\textwidth]{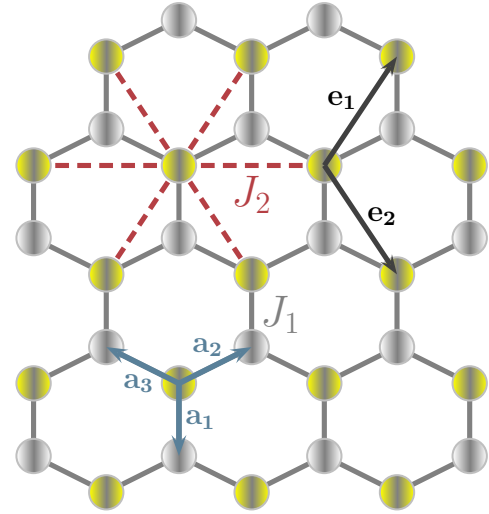}
   \caption{ (Color online) The honeycomb lattice with $J_1$ and $J_2$ couplings considered in this paper.
 The lattice sites with different colors belong to different sublattices.
 The primitive translation vectors of the direct lattice are $\left[ \mathbf{e}_{1}\text{=}%
\left( \sqrt{3}/2,3/2\right) \text{, }\mathbf{e}_{2}\text{=}\left( \sqrt{3}%
/2,-3/2\right) \right] $. $\mathbf{a}_{1}\text{=}\left( 0,-1\right) \text{, }\mathbf{a}_{2}%
\text{=}\left( \sqrt{3}/2,1/2\right) \text{ and }\mathbf{a}_{3}\text{=}\left( -%
\sqrt{3}/2,1/2\right)$ correspond to the nearest neighbor
bonds.} \label{fig:red}
\end{figure}

The study of frustrated quantum magnets on the honeycomb lattice has
also experimental motivations
\cite{BiMnO,ESR,expnew2,Azuma,libro_experimental,exp_hex2,expnew1}.
One of the most exciting experimental progresses is one kind of
bismuth oxynitrate, {\bimn}, which was obtained by Smirnova
\textit{et al.}\cite{BiMnO}. In this compound the Mn$^{4+}$ ions
form a $S=3/2$ honeycomb lattice without any distortion. The
magnetic susceptibility data indicates two-dimensional magnetism.
Despite the large AF Weiss constant of -257K, no long-range ordering
was observed down to 0.4K, which suggests a nonmagnetic ground
state\cite{BiMnO,ESR,expnew2,Azuma}. The substitution of Mn$^{4+}$
in {\bimn} by V$^{4+}$ may lead to the realization of the $S=1/2$
Heisenberg model on the honeycomb lattice.

The analysis of the honeycomb lattice from a more general point of
view has gained lately a lot of interest both coming from
graphene-related issues\cite{Neto} and from the possible spin-liquid
phase found in the Hubbard model in such geometry
\cite{HubbardNature,Wang,Lu,Yang,Vaezi1,Vaezi2,Tran,Sorella}. Due to
these reasons, recently there is huge theoretical interest in
frustrated Heisenberg models on the honeycomb lattice, in which
frustration is incorporated by second nearest neighbors
couplings\cite{Okumura,Wang,Mulder,Ganesh,Clark,Mosadeq,Cabra2012,Mezzacapo,Bishop}
and maybe also third nearest neighbors
couplings\cite{nos,Albuquerque,Oitmaa,Farnell,Reuther,Li_honeycomb_J1neg,Bishop_PRB,Li_2012_honeyJ1-J2-J3}.

Motivated by previous results, in this paper we study the
spin-1/2 Heisenberg model on the honeycomb lattice with first
($J_1$) and second ($J_2$) neighbors couplings. Using a Schwinger
boson mean field theory (SBMFT) we find strong evidence for the
existence of an intermediate disordered region where a spin gap
opens and spin-spin correlations decay exponentially. This
magnetically disordered region quantitatively agrees well with
recent numerical simulation
results\cite{Albuquerque,Mezzacapo,Bishop,Li_2012_honeyJ1-J2-J3}.
Another key finding of our work is the presence of two kinds of
magnetically disordered phases in this region. One is a gapped spin
liquid (GSL)\cite{Wen_1991,LM} with short-range N\'eel correlations,
maintaining the lattice translational and rotational symmetry. The
other phase is a staggered dimer valence-bond crystal (VBC), which is
also called lattice nematic\cite{Mulder}. This phase breaks lattice
rotational symmetry, but preserves lattice translational symmetry.

\begin{figure}[t!]
\includegraphics[width=0.45\textwidth]{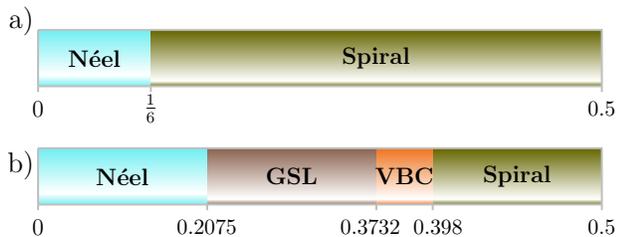}
   \caption{ (Color online) Phase diagram as a function of the frustration $J_{2}/J_{1}$. a) Classical phase diagram.
b) Quantum phase diagram corresponding to $S=\frac{1}{2}$ obtained
by means of SBMFT. } \label{fig:phase}
\end{figure}

The rest of the paper is arranged as follows. In Sec. II we
introduce our model and give a quick overview of the final phase
diagram. In Sec. III the general formalism of the Schwinger boson
mean-field approach is presented. In Sec. IV, using the solutions of
mean field equations, we discuss the phase diagram, especially the
magnetically disordered region. We close with a summary and
discussion in Sec. V.

\section{Model and overview of the phase diagram}

The $J_1-J_2$ Heisenberg model on the  honeycomb lattice is given by
\small
\ba
 \label{eq:Hspin_general}
 H =J_1\sum_{\langle \ve{x}\ve{y}\rangle_1} \hat{\bf{S}}_{\ve{x}}\cdot
\hat{\bf{S}}_{\ve{y}} +J_2\sum_{\langle \ve{x}\ve{y}\rangle_{2}} \hat{\bf{S}}_{\ve{x}}\cdot
\hat{\bf{S}}_{\ve{y}},
\ea
\normalsize
where $\hat{\bf{S}}_{\ve{x}}$ is the spin operator on site $\ve{x}$
and $\langle \ve{x}\ve{y}\rangle_n$ indicates sum over the $n$-th
neighbors (see Fig.~\ref{fig:red}). In this paper we are interested
in the antiferromagnetic case ($J_{1},J_{2}\geq 0$), and we focus on
the region $J_{2}/J_{1}\in \left[ 0,0.5\right] $.
\begin{figure}[t!]
\includegraphics[width=0.35\textwidth]{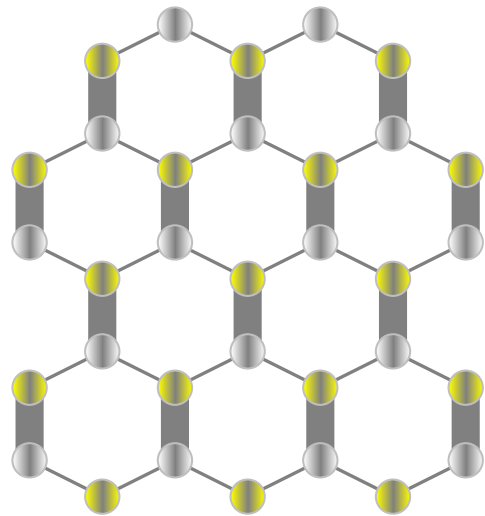}
   \caption{ (Color online) Sketch of the staggered dimer VBC state which breaks the
lattice rotational symmetry but preserves the lattice translational symmetry.
} \label{fig:VBC_sketch}
\end{figure}
In the classical limit, $S\to\infty$, the model displays different
zero temperature phases\cite{Katsura,Rastelli,Fouet}, see
Fig.~\ref{fig:phase}(a). For $J_2/J_1<1/6$, the system is N\'eel
ordered, while for $J_2/J_1>1/6$, the system shows spiral phases.
For the quantum case, aspects of this model have been explored
previously in the literature by various approaches, including spin
wave theory\cite{Rastelli,Fouet,Mulder,Ganesh}, non-linear
$\sigma$-model approach\cite{Einarsson}, mean field
theory\cite{Mattsson,Wang}, exact diagonalization
(ED)\cite{Fouet,Mosadeq,Albuquerque}, variational Monte Carlo (VMC)
method\cite{Clark,Mezzacapo}, series expansion (SE)\cite{Oitmaa},
pseudofermion functional renormalization group (PFFRG)\cite{Reuther}
and coupled cluster method (CCM)\cite{Bishop}. However, these works
yielded conflicting physical scenarios.

This model was studied by Mattsson \textit{et al.}\cite{Mattsson}
using SBMFT with a mean field decoupling that considers only
antiferromagnetic correlations for nearest neighbors and
ferromagnetic correlations for next nearest neighbors. This scheme
can only correctly describe N\'eel order. More recently
Wang\cite{Wang} studied this model within SBMFT including
antiferromagnetic correlations for both nearest and next nearest
neighbors. Unfortunately,  The author did not give the phase diagram
for different values of $J_{2}/J_{1}$. Actually, for frustrated
models we can not generally exclude either ferromagnetic or
antiferromagnetic correlations\cite{Coleman} and is important to use
a mean field decomposition that allows to include ferromagnetic an
antiferromagnetic correlations in equal footing. Another point is
that both of them assume the bond mean fields are independent of the
directions of bonds. Therefore, these two schemes can not describe
the phases in which the lattice rotational symmetry has broken. Here
we study the Hamiltonian (\ref{eq:Hspin_general}) in the strong
quantum limit($S=1/2$) using a rotationally invariant version of
this technique, which has proven successful in incorporating quantum
fluctuations\cite{Trumper1,Gazza,Trumper2,Coleman,nos,Mezio,Feldner,Messio}.

%
\begin{figure}[t!]
\includegraphics[width=0.5\textwidth]{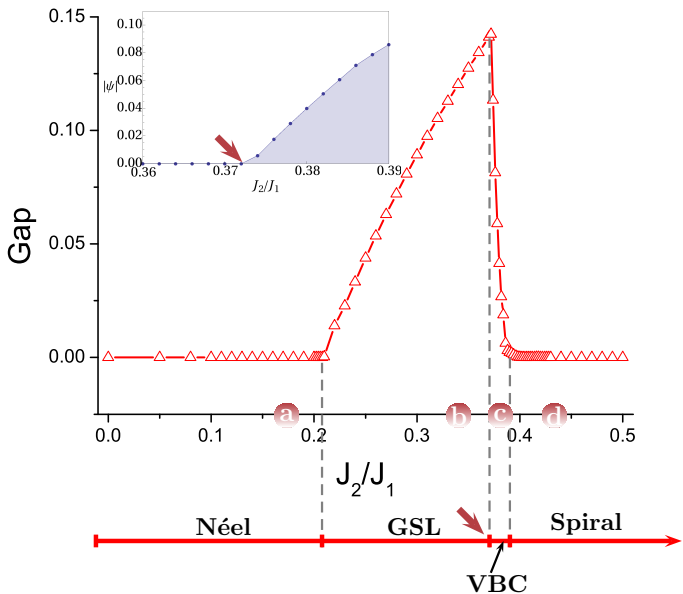}
   \caption{ (Color online) Gap in the boson dispersion extrapolated to the
thermodynamic limit as a function of the frustration $J_{2}/J_{1}$
corresponding to $S=1/2$. The gapped region corresponds to two
different magnetically disordered phases: one is GSL, the other is
staggered dimer VBC. Inset: $Z_3$ order parameter defined in Eq.
(\ref{eq:Z3}). The onset of the VBC phase is determined by the value
of
 $J_{2}/J_{1}$ where $|\psi|$ is non-zero (red arrows)} \label{fig:gap_and_phase_diag}
\end{figure}
%

Our main results are summarized in Fig.~\ref{fig:phase}(b). The
magnetic phase diagram is divided into four regions.\cite{error} At
small values of the frustrating coupling $J_{2}/J_{1}$, the system
presents a N\'eel-like ground state. By increasing the frustration,
we find at $J_2/J1 \simeq 0.2075$ a continuous transition
 to a gapped spin liquid phase.
When the value of the frustrating coupling exceeds $J_2/J1 \simeq
0.3732$,
 we find a continuous transition into a staggered dimer VBC (lattice nematic) with broken $Z_3$ symmetry
 (See Fig.~\ref{fig:VBC_sketch}),  which transforms at $J_{2}/J_{1} \simeq 0.398$ into a spiral
 phase.

\section{Schwinger boson mean-field approach}
\label{chap:SB}

It is well known that the SBMFT provides a natural description for
both magnetically ordered and disordered phases based on the picture
of the resonating valence bond
states\cite{Anderson,Auerbach,Auerbach:1994,Auerbach:2011}. As a
merit, this method does not start from any magnetic long range order
for the ground state (in contrast to spin wave theory), which should
emerge naturally if the Schwinger bosons condense at some momentum
vector\cite{Hirsch}. At this momentum vector, the lowest excitation spectrum of the
Schwinger bosons should be gapless. On the other hand, If the
Schwinger bosons are gapped, the phase is magnetically disordered.
 In the following, we will present in detail the rotationally
invariant version of SBMFT which was introduced by Ceccatto
\textit{et al.}\cite{Trumper1,Gazza,Trumper2} and we use in the following sections.

%
\begin{figure}[t!]
\includegraphics[width=0.5\textwidth]{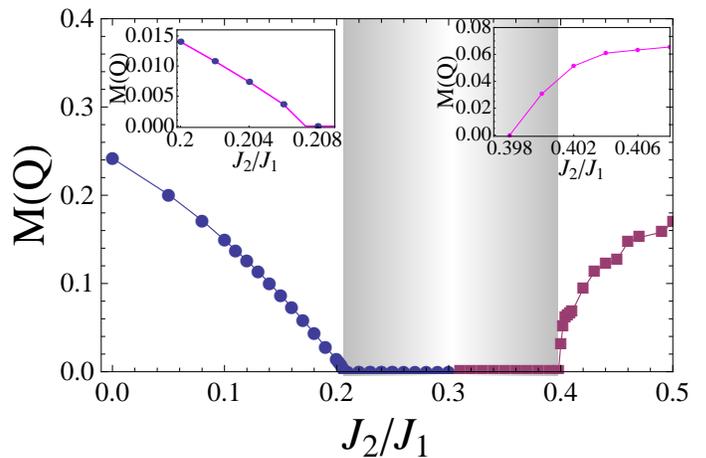}
   \caption{ (Color online) Local magnetization determined by  Eq. (\ref{eq:M(Q)}) extrapolated to the
thermodynamic limit as a function of the frustration $J_2/J_1$. The
shaded region corresponds to the magnetically disordered phases.
 Insets correspond to the regions where the magnetization for N\'eel (left) and Spiral (right) phases becomes zero.}
 \label{fig:mag}
\end{figure}
%

Consider the SU(2) Heisenberg Hamiltonian on a general lattice:
\ba
 \label{eq:H_gral_sb} \op{H}=\frac12
\sum_{\ve{x}\ve{y}\alpha\beta} J_{\alpha
\beta}(\ve{x}-\ve{y})\op{\ve{S}}_{\ve{x}+\ve{r}_{\alpha}} \cdot
\op{\ve{S}}_{\ve{y}+\ve{r}_{\beta}},
\ea
where $\ve{x}$ and $\ve{y}$ are the positions of the unit cells and vectors
$\ve{r}_{\alpha}$ are the positions of  each atom within the unit cell.
$J_{\alpha \beta}(\ve{x}-\ve{y})$ is the exchange interaction between the spins located in
 $\ve{x}+\ve{r}_{\alpha}$ and $\ve{y}+\ve{r}_{\beta}$.

In what follows we assume that the classical order can be parameterized as
\ba
\op{S}^{x}_{\ve{x}+\ve{r}_{\alpha}}&=&S\sin\varphi_{\alpha}(\ve{x})\\
\op{S}^{y}_{\ve{x}+\ve{r}_{\alpha}}&=&0\\
\op{S}^{z}_{\ve{x}+\ve{r}_{\alpha}}&=&S\cos\varphi_{\alpha}(\ve{x}),
\ea
with $\varphi_{\alpha}(\ve{x})={\bf Q}\cdot\ve{x}+\theta_{\alpha}$, where  ${\bf Q}$ is the ordering vector and
  $\theta_{\alpha}$ are the relative angles between the classical spins inside each unit cell.

The spin operators  $\op{\ve{S}}_\ve{x}$ on site $\ve{x}$ are represented by two bosons  $\op{b}_{\ve{x}\sigma}$
($\sigma=\uparrow,\downarrow$)
\ba
\label{eq:SBrepresentation}
 \op{\ve{S}}_\ve{r}=\frac12\; \op{\ve{b}}_\ve{r}^{\dag} \cdot
\vec{\sigma}\cdot \op{\ve{b}}_\ve{r}\, , \quad \op{\ve{b}}_\ve{r}=\left(\begin{array}{c}
  \op{b}_{\ve{r}\uparrow} \\
    \op{b}_{\ve{r}\downarrow}
     \end{array}\right),
\ea
where $\vec{\sigma}=(\sigma_x,\sigma_y,\sigma_z)$ are the Pauli matrices.
Eq. (\ref{eq:SBrepresentation}) is a faithful representation of the algebra SU(2) if we take into account the following local
constraint
\ba
\label{eq:sb constraint}
2S&=&\op{b}^{\dag}_{{\bf x}\,\uparrow} \op{b}_{{\bf x}\,\uparrow}+
\op{b}^{\dag}_{{\bf x}\,\downarrow}\op{b}_{{\bf x}\,\downarrow}.
 \ea
The exchange term can be expressed as

\ba
 \op{\ve{S}}_{\ve{x}+\ve{r}_{\alpha}} \! \cdot \!
\op{\ve{S}}_{\ve{y}+\ve{r}_{\beta}}&\!\!\!=&\!\!\! : \! \op{B}^{\dag}_{\alpha\beta}(\ve{x},\ve{y})
 \op{B}_{\alpha\beta}(\ve{x},\ve{y})\! :-
\op{A}^{\dag}_{\alpha\beta}(\ve{x},\ve{y})
\op{A}_{\alpha\beta}(\ve{x},\ve{y}),\nonumber\\
\ea
where $\op{A}_{\alpha,\beta}(\ve{x},\ve{y})$ and
$\op{B}_{\alpha,\beta}(\ve{x},\ve{y})$ are SU(2) invariants
defined as
\ba
\label{eq:A}
 \op{A}_{\alpha,\beta}(\ve{x},\ve{y})&=&\frac12 \sum_{\sigma}
\sigma \op{b}^{(\alpha)}_{\ve{x},\sigma}
\op{b}^{(\beta)}_{\ve{y},-\sigma}\\
\label{eq:B}
\op{B}_{\alpha,\beta}(\ve{x},\ve{y})&=&\frac12 \sum_{\sigma}
\op{b}^{\dag(\alpha)}_{\ve{x},\sigma}
\op{b}^{(\beta)}_{\ve{y},\sigma},
\ea
with $\sigma=\uparrow, \downarrow$. The double dots ($:\op{O}:$)
indicate the normal ordering of operator $\op{O}$.
This decoupling is particularly useful in the study of magnetic
systems near disordered phases, because it allows to treat
antiferromagnetism and ferromagnetism in equal
footing\cite{Trumper1,Trumper2,Gazza,Coleman}. On the other hand,
this scheme has been tested to obtain quantitatively quite accurate
results which show excellent agreements with
ED\cite{Trumper1,Gazza,Trumper2,nos}.

To construct a mean field Hamiltonian we perform the following Hartree-Fock decoupling
\small
\ba
 \nonumber (\op{\ve{S}}_{\ve{x}+\ve{r}_{\alpha}}\cdot
\op{\ve{S}}_{\ve{y}+\ve{r}_{\beta}})_{MF}&=& [B_{\alpha
\beta}^{*}(\ve{x}-\ve{y}) \op{B}_{\alpha \beta}(\ve{x},\ve{y})\\\nonumber
&-&
A_{\alpha \beta}^{*}(\ve{x}-\ve{y}) \op{A}_{\alpha \beta}(\ve{x},\ve{y})+H.c]\\
&-& \langle (\op{\ve{S}}_{\ve{x}+\ve{r}_{\alpha}}\cdot
\op{\ve{S}}_{\ve{y}+\ve{r}_{\beta}})_{MF} \rangle,
 \ea
\normalsize
where we have defined
\small
 \ba
 \label{eq:MF_eq_SB1}
 A_{\alpha \beta}^{*}(\ve{x}-\ve{y})&=&\langle  \op{A}_{\alpha \beta}^{\dag}(\ve{x},\ve{y})\rangle\\
 \label{eq:MF_eq_SB2}
 B_{\alpha \beta}^{*}(\ve{x}-\ve{y})&=&\langle  \op{B}_{\alpha \beta}^{\dag}(\ve{x},\ve{y})\rangle\\\nonumber
 \label{eq:MF_eq_SB3}
\langle (\op{\ve{S}}_{\vx+\vra}\!\!\cdot \!\! \op{\ve{S}}_{\vy+\vrb} )_{MF} \rangle
\! &=&\! |B_{\alpha \beta}(\vx-\vy)|^{2}-|A_{\alpha
\beta}(\vx-\vy)|^{2},
 \ea
\normalsize
and $\langle\;\rangle$ denotes the expectation value in the ground state at $T=0$.
It is convenient to change variables to $\ve{R}=\ve{x}-\ve{y}$,
and eliminating  $\ve{x}$ in the sums we obtain
 %
\small
\ba
\nonumber
\op{H}_{MF}&=& \frac12 \sum_{\ve{R}\ve{y}\alpha\beta} J_{\alpha \beta}(\ve{R})
\left\{\frac12 \sum_{\sigma}\left[
B_{\alpha,\beta}(\ve{R})\;\op{b}^{\dag(\alpha)}_{\ve{R}+\ve{y},\sigma}
\op{b}^{(\beta)}_{\ve{y},\sigma}
\right.\right.\\\nonumber
&-&\left.\sigma
A_{\alpha,\beta}(\ve{R})\;\op{b}^{\dag
(\alpha)}_{\ve{R}+\ve{y},\sigma} \op{b}^{\dag
(\beta)}_{\ve{y},-\sigma} +H.C.\right]\\\nonumber
 &-&\left.
\left(\;|B_{\alpha,\beta}(\ve{R})|^{2}-|A_{\alpha,\beta}(\ve{R})|^{2}\right)\right\}.
%
\ea
%
%
\normalsize

The mean field Hamiltonian is quadratic in the boson operators and
can be diagonalized. It is convenient to transform the
operators to momentum space
\ba
 \op{b}^{(\alpha)}_{\ve{x},\sigma}=\frac{1}{\sqrt{N_{c}}}
\sum_{\ve{k}}\op{b}^{(\alpha)}_{\ve{k},\sigma}e^{i\ve{k} \cdot
(\ve{x}+\ve{r}_{\alpha})} ,\ea
where $N_{c}$ is the number of unit cells. After some algebra and
using the symmetry properties:
 \ba
\label{eq:simetrias1}
\nonumber  J_{\alpha \beta}(\ve{R})&=&J_{\beta \alpha }(-\ve{R})\\
  A_{\alpha \beta}(\ve{R})&=&-A_{\beta \alpha }(-\ve{R})\\
\nonumber  B_{\alpha \beta}(\ve{R})&=&B^{*}_{\beta \alpha }(-\ve{R}),
 \ea
we obtain the following form for the Hamiltonian
\begin{widetext}
\small
 \ba
\nonumber
 \op{H}_{MF}&=&\frac12
 \sum_{\ve{k}\alpha\beta}\sum_{\sigma}\left\{ \gamma^{B}_{\alpha\beta}(\ve{k})\op{b}^{\dag(\alpha)}_{\ve{k}\sigma}
 \op{b}^{(\beta)}_{\ve{k}\sigma}+
 \gamma^{B}_{\alpha\beta}(-\ve{k})\op{b}^{\dag(\alpha)}_{-\ve{k}-\sigma}
 \op{b}^{(\beta)}_{-\ve{k}-\sigma}
  - \sigma \gamma^{A}_{\alpha\beta}(\ve{k})\op{b}^{\dag(\alpha)}_{\ve{k}\sigma}
 \op{b}^{\dag(\beta)}_{-\ve{k}-\sigma}
 -\sigma  \bar{\gamma}^{A}_{\alpha\beta}(\ve{k})\op{b}^{(\alpha)}_{\ve{k}\sigma}
 \op{b}^{(\beta)}_{-\ve{k}-\sigma}
 \right\}\\
 &&-\frac{N_{c}}{2}\sum_{\ve{R}\alpha\beta}J_{\alpha\beta}(\ve{R})\left[|B_{\alpha\beta}(\ve{R})|^{2}-
 |A_{\alpha\beta}(\ve{R})|^{2}
 \right],
 \ea
\end{widetext}
\normalsize
where
\small
 \ba
 \!\!\gamma^{B}_{\alpha \beta}(\ve{k})\!\!&=&\!\!\frac12 \sum_{\ve{R}}\!
 J_{\alpha\beta}(\ve{R}) B_{\alpha\beta}(\ve{R}) e^{-i \ve{k}\cdot
 (\ve{R}+\ve{r}_{\alpha}-\ve{r}_{\beta})}\\
 \!\!\gamma^{A}_{\alpha \beta}(\ve{k})\!\!&=&\!\!\frac12 \sum_{\ve{R}}\!
 J_{\alpha\beta}(\ve{R}) A_{\alpha\beta}(\ve{R}) e^{-i \ve{k}\cdot
 (\ve{R}+\ve{r}_{\alpha}-\ve{r}_{\beta})}\\
 \! \!\bar{\gamma}^{A}_{\alpha \beta}(\ve{k})\!\!&=&\!\!\frac12 \sum_{\ve{R}}\!
 J_{\alpha\beta}(\ve{R}) \bar{A}_{\alpha\beta}(\ve{R}) e^{-i \ve{k}\cdot
 (\ve{R}+\ve{r}_{\alpha}-\ve{r}_{\beta})}.
 \ea
\normalsize

Now, we impose the constraint (\ref{eq:sb constraint}) in average over each
sublattice $\alpha$ by means of  Lagrange multipliers
$\lambda^{(\alpha)}$
 \ba
 \op{H}_{MF}\rightarrow
 \op{H}_{MF}+\op{H}_{\lambda}
 \ea
with
 \ba
 \op{H}_{\lambda}=\sum_{\ve{x}\alpha}\lambda^{(\alpha)}
 \left(\sum_{\sigma}\op{b}^{\dag(\alpha)}_{\ve{x}\sigma}\op{b}^{(\alpha)}_{\ve{x}\sigma}-2S\right).
 \ea

Using the symmetries (\ref{eq:simetrias1}) we can see that both kinds of
bosons ($\uparrow,\downarrow$)  give the same contribution to the Hamiltonian.
Then, we can perform the sum over  $\sigma$ to obtain
\begin{widetext}
 \ba
\nonumber \op{H}_{MF}&=&\frac12 \sum_{\ve{k}\alpha \beta}\left\{ (
\gamma^{B}_{\alpha \beta}(\ve{k})+\lambda^{(\alpha)}\delta_{\alpha
\beta})
\op{b}^{\dag(\alpha)}_{\ve{k}\uparrow}\op{b}^{(\beta)}_{\ve{k}\uparrow}+
 ( \gamma^{B}_{\alpha \beta}(-\ve{k})+\lambda^{(\alpha)}\delta_{\alpha \beta})
\op{b}^{\dag(\alpha)}_{-\ve{k}\downarrow}\op{b}^{(\beta)}_{-\ve{k}\downarrow}
 -
 \sigma \left(\gamma^{A}_{\alpha \beta}(\ve{k})\op{b}^{\dag(\alpha)}_{\ve{k}\uparrow}\op{b}^{\dag(\beta)}_{-\ve{k}\downarrow}
+  \bar{\gamma}^{A}_{\alpha
\beta}(\ve{k})\op{b}^{(\alpha)}_{\ve{k}\uparrow}\op{b}^{(\beta)}_{-\ve{k}\downarrow}
\right)
\right\}\\
\nonumber
&&-\frac{N_c}{2}\sum_{\ve{R}\alpha\beta}J_{\alpha\beta}(\ve{R})\left[|B_{\alpha\beta}(\ve{R})|^{2}-
 |A_{\alpha\beta}(\ve{R})|^{2}
 \right]-2SN_{c}\sum_{\alpha}\lambda^{(\alpha)}.
 \ea
\end{widetext}

It is convenient to introduce the Nambu spinor
$\op{\ve{b}}^{\dag}(\ve{k})=\left(
\hat{\ve{b}}^{\dag}_{\ve{k}\uparrow},\hat{\ve{b}}_{-\ve{k}\downarrow}
\right)$ where
 \ba
  \hat{\ve{b}}^{\dag}_{\ve{k}\uparrow}&=& (\op{b}^{\dag(\alpha_{1})}_{\ve{k}\uparrow},\op{b}^{\dag(\alpha_{2})}_{\ve{k}\uparrow},...,
\op{b}^{\dag(\alpha_{n_{c}})}_{\ve{k}\uparrow})\\
\hat{\ve{b}}_{-\ve{k}\downarrow}&=&
(\op{b}^{\dag(\alpha_{1})}_{-\ve{k}\downarrow},\op{b}^{\dag(\alpha_{2})}_{-\ve{k}\downarrow},...,
\op{b}^{\dag(\alpha_{n_{c}})}_{-\ve{k}\downarrow})
 \ea
 and  $n_{c}$ is the number of atoms in the unit cell.
Now, we can rewrite the Hamiltonian into a compact form:
\ba
 \label{eq:HMF_compacto}
 H_{MF}&=&\sum_{\vk}\; \op{\ve{b}}^{\dag}(\ve{k})\cdot D(\vk) \cdot
\op{\ve{b}}(\ve{k})\\\nonumber
&-&(2S+1)N_{c}\sum_{\alpha}\lambda^{(\alpha)}- \langle H_{MF} \rangle,
\ea
where the   $2\, n_c \times 2\, n_c $ dynamical matrix $D(\ve{k})$  is given by
\small
 \ba
\nonumber
 D(\ve{k})\!=\!\left(
  \begin{tabular}{cc}
  $\gamma^{B}_{\alpha \beta}(\ve{k})+\lambda^{(\alpha)}\delta_{\alpha \beta}$ & $-\gamma^{A}_{\alpha \beta}(\ve{k})$ \\
$\gamma^{A}_{\alpha \beta}(\ve{k})$ & $\gamma^{B}_{\alpha
\beta}(\ve{k})+\lambda^{(\alpha)}\delta_{\alpha \beta}$
  \end{tabular}
\right).
 \ea
\normalsize
%
%
%

To diagonalize the Hamiltonian (\ref{eq:HMF_compacto}) we need to
perform a para-unitary transformation of the matrix  $D(\ve{k})$
which preserves the bosonic commutation relations\cite{Colpa}. We
can diagonalize the Hamiltonian by defining the new operators
$\op{\ve{a}}=F \cdot \op{\ve{b}}$, where the matrix $F$ satisfy
\be
(F^{\dag})^{-1}\cdot \tau_3 \cdot (F)^{-1}= \tau_3 ,\quad \tau_3=
\left(
\begin{array}{cc}
I_{2\times 2} & 0\\
0 & -I_{2\times 2}
\end{array}
\right).
\ee
With this transformation, the Hamiltonian reads
\ba
\op{H}_{MF}=\sum_{\ve{k}} \op{\ve{a}}^{\dag}_{\ve{k}} \cdot \ve{E}(\ve{k}) \cdot \op{\ve{a}}_{\ve{k}}- (2S+1)N_{c}\sum_{\alpha}\lambda^{(\alpha)}-\langle \op{H}_{MF} \rangle, \nonumber\\
\ea
where
\ba
\ve{E}(\ve{k})=\mbox{diag}(\omega_{1}(\ve{k}),...,\omega_{n_{c}}(\ve{k}),\omega_{1}(\ve{k}),
...,\omega_{n_{c}}(\ve{k})). \ea
%

In terms of the original bosonic operators, the mean field
parameters are
\ba
\nonumber
A_{\alpha \beta}(\ve{R})\!\! &=&\!\! \frac{1}{2 N_{c}}\sum_{\ve{k}}
\left\{
e^{i\ve{k}(\ve{R}+\ve{r}_{\alpha}-\ve{r}_{\beta})} \langle
  \op{b}^{(\alpha)}_{\ve{k}\uparrow} \op{b}^{(\beta)}_{-\ve{k}\downarrow}\rangle \right.\\
\label{eq:sc1}
&-&\left. e^{-i\ve{k}(\ve{R}+\ve{r}_{\alpha}-\ve{r}_{\beta})} \langle  \op{b}^{(\alpha)}_{-\ve{k}\downarrow} \op{b}^{(\beta)}_{\ve{k}\uparrow}\rangle
 \right\} \\\nonumber
B_{\alpha \beta}(\ve{R})\!\! &=&\!\! \frac{1}{2 N_{c}}\sum_{\ve{k}}
\left\{
e^{i\ve{k}(\ve{R}+\ve{r}_{\alpha}-\ve{r}_{\beta})}
 \langle  \op{b}^{\dag(\beta)}_{\ve{k}\uparrow} \op{b}^{(\alpha)}_{\ve{k}\uparrow}\rangle  \right.\\
\label{eq:sc2}
&-&\left. e^{-i\ve{k}(\ve{R}+\ve{r}_{\alpha}-\ve{r}_{\beta})} \langle  \op{b}^{\dag (\beta)}_{-\ve{k}\downarrow} \op{b}^{(\alpha)}_{-\ve{k}\downarrow}\rangle
 \right\}
\ea
and the constraint in the number of bosons can be written in the momentum space as
%
\small
\ba
\label{eq:const1}
\sum_{\ve{k}} \left\{
 \langle  \op{b}^{\dag(\alpha)}_{\ve{k}\uparrow} \op{b}^{(\alpha)}_{\ve{k}\uparrow} \rangle +
\langle  \op{b}^{\dag(\alpha)}_{-\ve{k}\downarrow} \op{b}^{(\alpha)}_{-\ve{k}\downarrow} \rangle
 \right\}=2S N_{c},
\ea
\normalsize
where $N_c$ is the total number of unit cells and $S$ is the spin strength.
The mean field equations (\ref{eq:sc1}) and (\ref{eq:sc2}) must be solved in a self-consistent way together with the
constraints (\ref{eq:const1}) on the number of bosons.

Finding numerical solutions involves finding the roots of the
coupled nonlinear equations for the parameters $A$ and $B$, plus the
additional constraints to determine the values of the Lagrange
multipliers $\lambda^{(\alpha)}$. We perform the calculations for
finite but very large lattices and  finally we extrapolate the
results to the thermodynamic limit.

We solve numerically for different values of the frustration
parameter $J_2/J_1$ and with the values obtained for the MF
parameters and the Lagrange multipliers we compute the energy and
the new values for the MF parameters.
We repeat this self-consistent procedure until the energy and the MF
parameters converge. After reaching convergence we can compute all
physical quantities like the energy, the excitation gap, the
spin-spin correlation and the local magnetization.  During the
calculation, it is convenient to fix the energy scale by setting the
value of the nearest-neighbor coupling $J_{1}=1$.

\section{Results}

In Fig.~\ref{fig:gap_and_phase_diag}, we show the boson dispersion
relation gap extrapolated to the thermodynamic limit as a function of the frustration ($J_{2}/J_{1}$). In the gapped region,
the absence of Bose condensation indicates that the ground state is
magnetically disordered. This result agrees well with recent
ED\cite{Albuquerque}, VMC\cite{Mezzacapo} and
CCM\cite{Bishop,Li_2012_honeyJ1-J2-J3} studies. In the gapless
region, the excitation spectrum is zero at a given wave vector $%
\mathbf{k}^{\ast }=\mathbf{Q/}2$, where the Boson condensation
occurs. This is characteristic of  the magnetically ordered phases. The structure
of these phases can be understood through the spin-spin correlation
function (SSCF) and the excitation spectrum. Some typical examples
for different phases will be shown later.

\begin{figure}[t!]
\includegraphics[width=0.47\textwidth]{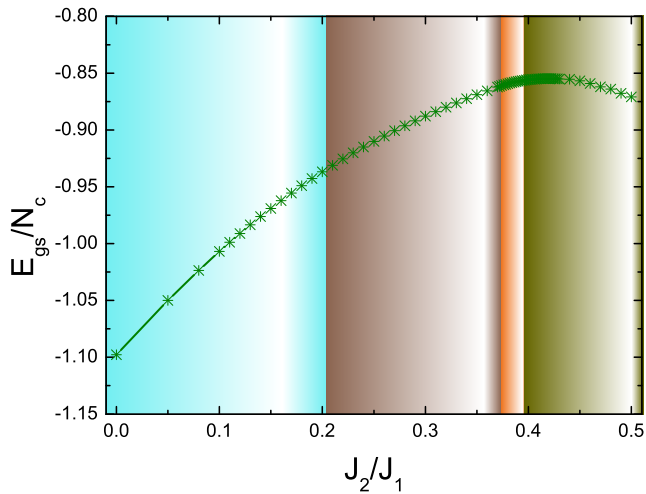}
   \caption{ (Color online) Ground-state energy per unit cell extrapolated to the
thermodynamic limit as a function of the frustration $J_{2}/J_{1}$.
 The regions of the four different phases are indicated using the same colors that are used in Fig.~\ref{fig:phase}.} \label{fig:Egs}
\end{figure}

To pin down the precise phase boundaries between the magnetically
ordered and disordered phases, we introduce the local magnetization
$M(\mathbf{Q)}$ as an order parameter, which is obtained from the
long distance behavior of the spin-spin correlation function
(SSCF)\cite{Trumper1,Gazza}:
\be
\label{eq:M(Q)}
\lim_{\left\vert \mathbf{x-y}\right\vert \rightarrow \infty
}\left\langle \mathbf{S}_{\mathbf{x}}\cdot
\mathbf{S}_{\mathbf{y}}\right\rangle \approx
M^{2}\left( \mathbf{Q}\right) \cos \left[ \mathbf{Q\cdot }\left( \mathbf{x-y}%
\right) \right].
 \ee
 In Fig.~\ref{fig:mag}, we show the local
magnetization for $J_{2}/J_{1}\in \left[ 0,0.5\right] $. For
$J_2/J_1=0$, the local magnetization is $M(\mathbf{Q)=}0.24176$,
which is in excellent agreement with the second order spin wave
calculation result of 0.2418.\cite{Zheng} This value is
significantly reduced by quantum fluctuations compared with the
classical value $0.5$. The quantum Monte Carlo (QMC)
result\cite{Castro} is $0.2677(6)$, which is considerably larger
than ours. For the unfrustrated case, all the mean field approaches
are quite inaccurate compared with much more controlled techniques
like QMC. The difference in the $M(\textbf{Q})$ values of about
10{\%}, provides, in the absence of any other quantitative evidence
for the accuracy of the method as applied to this model, an
indication of the accuracy of the method and of all the results
quoted that depend on the order parameters, including the phase
boundaries. However, the mean field approach is still very useful to
study gapped phases in frustrated systems. On one hand it is well
known that for frustrated systems QMC presents the famous sign
problem. On the other hand, the study of quantities like energy gap
requires the study of big sizes clusters and the use of exact
diagonalization for small size clusters makes it very difficult to
extrapolate the results.

As $J_2/J_1$ increases, the local magnetization decreases. It
vanishes continuously at $J_{2}/J_{1}\simeq 0.2075$, as shown in
Fig.~\ref{fig:mag}.\cite{error} This value is in excellent agreement
with recent numerical results, such as $0.2$ by Mezzacapo {\it et
al.}\cite{Mezzacapo} using VMC with an entangled-plaquette
variational ansatz, as well as $0.207\pm 0.003$ by Bishop {\it et
al.}\cite{Bishop} using CCM.
The shift of N\'eel boundary compared with the classical estimate
$1/6$ is due to quantum fluctuations which prefer to collinear
N\'eel rather than spiral phases in some cases.\cite{Albuquerque} In
this region, the SSCF is antiferromagnetic in all directions, and
the Boson condensation happens at the $\Gamma$ point of the first
Brillouin zone: $\mathbf{k}^{\ast }=(0,0)$, which corresponds to the
ordering vector $\mathbf{Q}=(0,0)$.
As $J_2/J_1$ decreases from 0.5, the local magnetization
$M(\mathbf{Q)}$ decreases. It vanishes continuously at
$J_{2}/J_{1}\simeq 0.398$, as shown in
Fig.~\ref{fig:mag}.\cite{error} This value is also in good agreement
with recent numerical results, such as $0.4$ by Mezzacapo {\it et
al.}\cite{Mezzacapo}, as well as $0.385\pm 0.010$ by Bishop {\it et
al.}\cite{Bishop}.
In this region, the SSCF shows different properties in different
directions, however, it exhibits long range order in all directions.
The gapless points of the excitation spectrum move continuously
inside the first Brillouin zone as $J_2/J_1$ changes. This results
correspond to a spiral phase. In the classical version
($S\to\infty$) of the model (See Fig.~\ref{fig:phase}(a)), for
$J_{2}/J_{1}>1/6$ there remains a line-type degeneracy in which the
spiral wave number is not determined uniquely and is allowed on a
ring in the Brillouin zone.\cite{Rastelli,Fouet} Our results suggest
that the classical degeneracy is lifted in the quantum version,
where some spiral wave vectors are favored by quantum fluctuations
from the manifold of classically degenerate spiral wave vectors.
This spiral order by disorder selection was already seen by using a
spin wave approach by Mulder {\it et al.},\cite{Mulder} and we have
recovered this selection with a different approach.

The most interesting part of the phase diagram is the intermediate region which has no
classical counterpart. In this region, the nonmagnetic ground state
retains SU(2) spin rotational symmetry and the lattice translational
symmetry, However, it may break the $Z_3$ directional symmetry of
the lattice. Following Mulder {\it et al.}\cite{Mulder}  we
introduce the $Z_3$ directional symmetry breaking order parameter
$\left\vert \psi \right\vert$ where
\begin{eqnarray}
\label{eq:Z3}
\psi  &=&\left\langle \mathbf{S}_{A}\left( \mathbf{r}\right) \cdot \mathbf{S}%
_{B}\left( \mathbf{r}\right) \right\rangle +\omega \left\langle \mathbf{S}%
_{A}\left( \mathbf{r}\right) \cdot \mathbf{S}_{B}\left( \mathbf{r+e}%
_{1}\right) \right\rangle  \nonumber\\
&&+\omega ^{2}\left\langle \mathbf{S}_{A}\left( \mathbf{r}\right)
\cdot \mathbf{S}_{B}\left( \mathbf{r-e}_{2}\right) \right\rangle.
\end{eqnarray}
Here $A$, $B$ correspond to the two different sublattices, $\mathbf{r}$
denotes the unit cell position, and $\omega =\exp \left( i2\pi /3\right) $.
Equivalently, Okumura {\it et al.}\cite{Okumura} define $\mathbf{m}_{3}=\varepsilon _{1}%
\mathbf{a}_{1}+\varepsilon _{2}\mathbf{a}_{2}+\varepsilon
_{3}\mathbf{a}_{3}$, where $\varepsilon _{\mu }$ $(\mu =1,2,3)$ are
bond energies corresponding to the three nearest neighbor bonds
$\mathbf{a}_{\mu }$ $(\mu =1,2,3)$. It is trivial to see $\left\vert
\psi \right\vert=\left\vert \mathbf{m}_{3} \right\vert$. This order
parameter is zero when the spin correlations along the three
directions are equal. We find that $\left\vert \psi \right\vert$
keeps zero when $J_{2}/J_{1}\lesssim 0.3732$; it becomes non-zero
continuously at $J_2/J1 \simeq 0.3732$ as shown in
Fig.~\ref{fig:gap_and_phase_diag}.\cite{error} Therefore, in the
region $0.2075\lesssim J_{2}/J_{1}\lesssim 0.3732$, the ground state
preserves the $Z_3$ lattice rotational symmetry. The SSCF shows
short range antiferromagnetic correlations in all directions, and
the minimum of the excitation spectrum remains pinned at the
$\Gamma$ point. Namely, the system remains to be a GSL. The
appearance of the GSL agrees with recent two different VMC
studies.\cite{Clark,Mezzacapo} In the region $0.3732\lesssim
J_{2}/J_{1}\lesssim 0.398$, the $Z_3$ lattice rotational symmetry
has broken. We find that the values of the mean fields $A$ and $B$:
$A(B)_{\mathbf{a}_{2}}=A(B)_{\mathbf{a}_{3}}\neq
A(B)_{\mathbf{a}_{1}}$; the bond energies have the same property:
$\varepsilon _{2}=\varepsilon _{3}\neq \varepsilon _{1}$. Therefore,
the system should belong to the staggered dimer VBC (lattice
nematic). To further analyze this region, one need to calculate the
dimer-dimer correlations. However, it is out of the scope of the
present paper. The existence of the staggered dimer VBC is in
agreement with a recent ED study,\cite{Mosadeq} a bond operator mean
field study,\cite{Mulder} and a VMC study.\cite{Clark}

\begin{figure}[t!]
\includegraphics[width=0.45\textwidth]{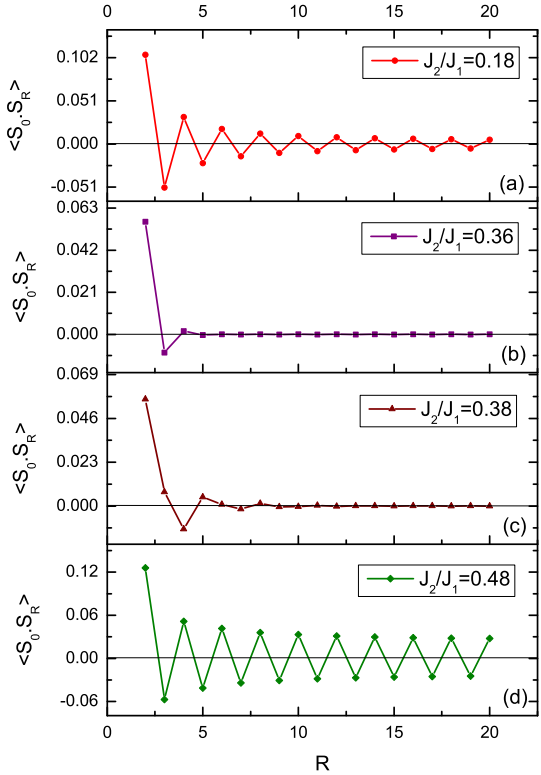}
   \caption{ (Color online) SSCF for a system of size $N=2\times 50\times 50$ in the zigzag direction corresponding to the four different phases: (a)$J_{2}/J_{1}=0.18$ (N\'eel), (b)$J_{2}/J_{1}=0.36$ (GSL),
    (c)$J_{2}/J_{1}=0.38$ (staggered dimer VBC), and (d)$J_{2}/J_{1}=0.48$ (spiral).} \label{fig:SSCF_zigzag}
\end{figure}
\begin{figure}[t!]
\includegraphics[width=0.45\textwidth]{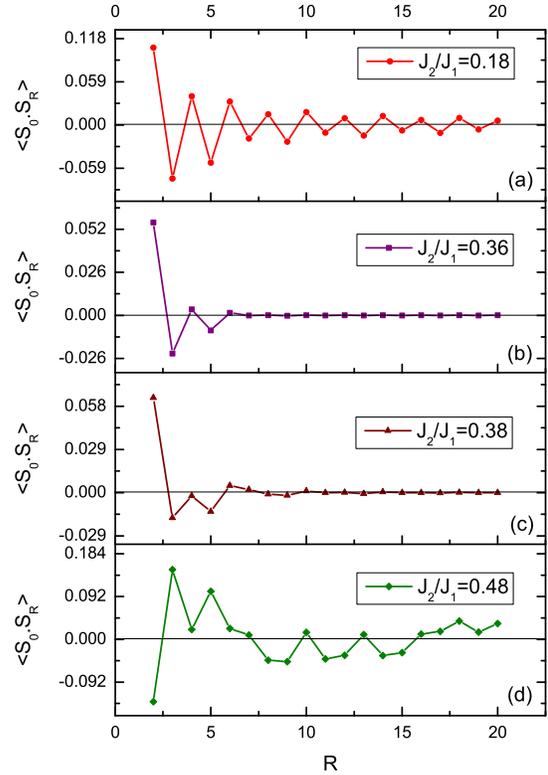}
   \caption{ (Color online) SSCF for a system of size $N=2\times 50\times 50$ in the armchair direction corresponding to the four different phases: (a)$J_{2}/J_{1}=0.18$ (N\'eel), (b)$J_{2}/J_{1}=0.36$ (GSL),
    (c)$J_{2}/J_{1}=0.38$ (staggered dimer VBC), and (d)$J_{2}/J_{1}=0.48$ (spiral).} \label{fig:SSCF_armchair}
\end{figure}

The errors in the values of the phase boundaries that are implicit
here in the number of significant figures quoted, correspond purely
to the error in the extrapolation of our finite-size results to the
thermodynamic limit. In no way are they intended to represent the
essentially unknown errors implicit in the mean-field approach,
e.g., the 10{\%} difference in $M(\textbf{Q})$ compared with the QMC
result in the unfrustrated limit. All the transition values
presented in this paper correspond to mean field estimations. In
order to improve these values, it is necessary to study in detail
the phase transitions beyond the mean field level, which is out of
the scope of the present paper.

In Fig.~\ref{fig:Egs} we show the results for the ground state
energy per unit cell extrapolated to the thermodynamic limit. For
the unfrustrated case ($J_{2}=0$), $E_{gs}/N_{c}$=-1.09779, which is
in excellent agreement with the second order spin wave calculation
result of $-1.0978$.\cite{Zheng} Compared with published QMC results
by Reger {\it et al.}\cite{Reger}: $-1.0890(9)$, and more recently
by L\"ow\cite{Low}: $-1.08909(39)$, it has appreciable difference,
as our previous discussion of the difference in the $M(\textbf{Q})$
values. Since energy estimates always have an intrinsic quadratic
error, compared to an intrinsic linear error for other properties,
even small errors in the energy can be of significance. The shape of
the energy curve also supports that the three quantum phase
transitions are continuous.

%
\begin{figure*}[!htb]
\includegraphics[width=0.85\textwidth]{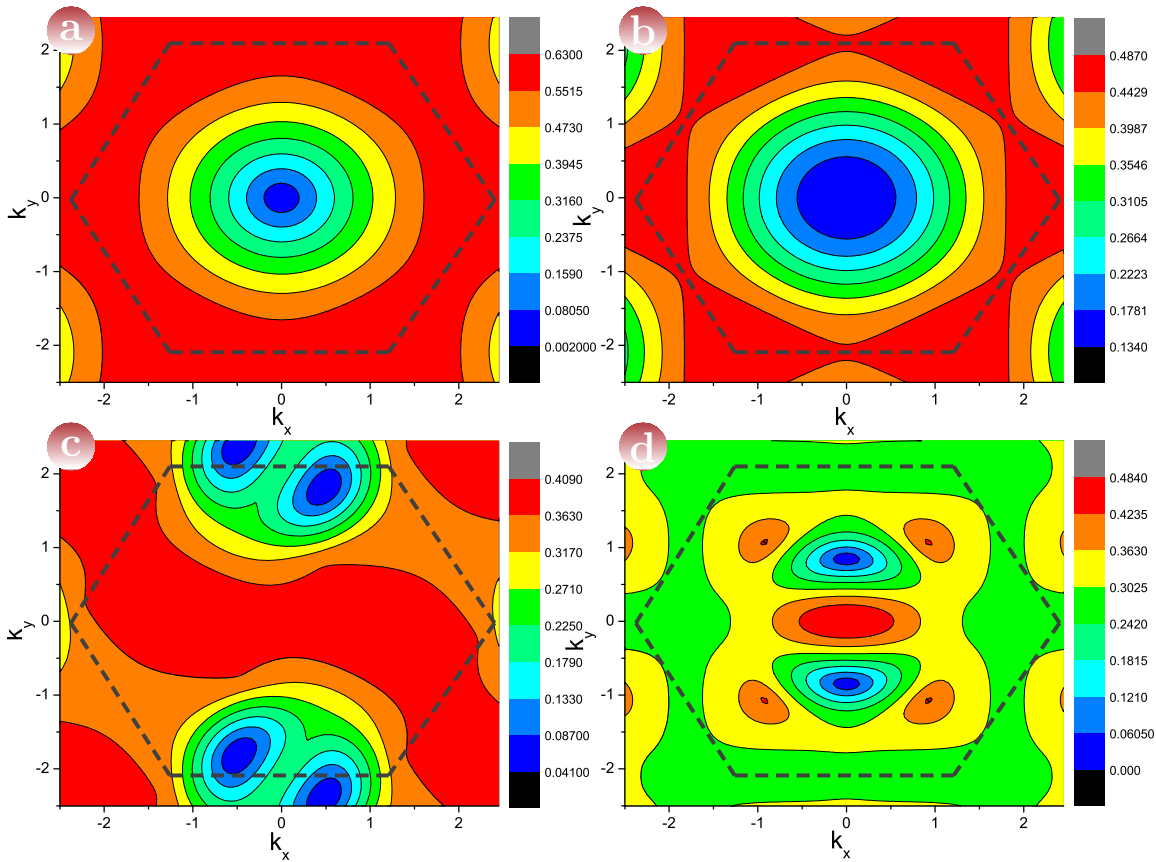}
   \caption{ (Color online) Momentum dependence of the lowest excitation spectrum for a system of size $N=2\times 50\times 50$ corresponding to the four different phases: (a)$J_{2}/J_{1}=0.18$ (N\'eel), (b)$J_{2}/J_{1}=0.36$ (GSL),
    (c)$J_{2}/J_{1}=0.38$ (staggered dimer VBC), and (d)$J_{2}/J_{1}=0.48$ (spiral). The dashed hexagon denotes the first Brillouin zone of the lattice.}
 \label{fig:spectrum}
\end{figure*}
%

In the following we show several typical examples for the four
different phases. The SSCF along zigzag and armchair directions for
a system of 5000 sites is shown in Fig.~\ref{fig:SSCF_zigzag} and
Fig.~\ref{fig:SSCF_armchair} for $J_{2}/J_{1}=0.18$ (N\'eel), 0.36
(GSL), 0.38 (staggered dimer VBC) and 0.48 (spiral). The
corresponding lowest excitation spectrum is shown in
Fig.~\ref{fig:spectrum}. Although it is a finite size system, we can
still see the corresponding properties for the four
different phases as we have presented above. For $J_{2}/J_{1}=0.18$,
the SSCF in both of the zigzag and armchair directions shows long
range N\'eel correlations, and the lowest excitation spectrum
becomes gapless at the $\Gamma$ point (for a finite size system there is
 a small gap which disappears after the extrapolation). For $J_{2}/J_{1}=0.36$, the
SSCF in both of the zigzag and armchair directions shows short range
N\'eel correlations, and the minimum of the lowest excitation
spectrum remains at the $\Gamma$ point, however, there is a large
gap which does not disappear after the extrapolation. For
$J_{2}/J_{1}=0.38$, the SSCF does not show any long range
correlation, and the short range correlations are different along
the zigzag or armchair directions, which is a indication that the
lattice rotational symmetry is broken. Simultaneously, the minimum
of the lowest excitation spectrum is away from the $\Gamma$ point
and the lattice rotational symmetry is clearly broken. There is also
a gap in this region which remains finite in the thermodynamic
limit. For $J_{2}/J_{1}=0.48$, the SSCF shows magnetic long range
correlations in both of the zigzag and armchair directions. Since
one component of the ordering vector $Q_{x}=0$ (corresponding to
$k_{x}^{\ast }=0$ in the lowest excitation spectrum), the SSCF is
N\'eel-like along the zigzag directions. This result agrees well
with the spin wave calculations by Mulder {\it et al.}.\cite{Mulder}

Finally, we would like to talk about the next step of our work. We
have used a mean field approach based in the Schwinger boson
representation of the spin operators. This mean field approach has
the drawback of being defined in a constrained bosonic space, with
unphysical configurations being allowed if this constraint is
treated as an average restriction. This drawback can be in principle
corrected by including local fluctuations of the bosonic chemical
potential.\cite{Raykin} This correction was calculated by Trumper
{\it et al.}\cite{Trumper2} for the $J_1-J_2$ square lattice using
collective coordinate methods, where a comparison between the mean
field results and the corrected results was made. However, this hard
calculation allows only to calculate some special quantities like
the ground state energy or spin stiffness. The corrections developed
by Trumper {\it et al.} could be extended to spiral
phases\cite{Manuel}, which would allow to investigate, for instance,
the present model.

\section{Summary and discussion}

In the present paper, we have investigated the quantum $J_1-J_2$
Heisenberg model on the honeycomb lattice within a rotationally
invariant version of SBMFT. In the region $J_{2}/J_{1}\in \left[
0,0.5\right] $, the quantum phase diagram of the model displays four
different regions.\cite{error} The magnetic long range order of
N\'eel and spiral types is found for $J_{2}/J_{1}\lesssim 0.2075$
and $J_{2}/J_{1}\gtrsim 0.398$, respectively. For the spiral region,
we get the spiral order from quantum disorder selection which agrees
with Mulder {\it et al.}\cite{Mulder} using spin wave theory. In the
intermediate region, the energy gap is finite while the local
magnetization is zero, which indicates the presence of a
magnetically disordered ground state.
 We have used
the $Z_3$ directional symmetry breaking order parameter $\left\vert
\psi \right\vert$ defined in Eq. (\ref{eq:Z3}) to classify this part
into two different magnetically disordered phases: one is a GSL
which shows short-range N\'eel correlations ($J_{2}/J_{1}\lesssim
0.3732$), the other is staggered dimer VBC (lattice nematic), which
breaks the $Z_3$ directional symmetry ($J_{2}/J_{1}\gtrsim 0.3732$).
Considering the properties of order parameters and the ground state
energy, these three quantum phase transitions seem to be continuous.

As we have mentioned above, recent theoretical studies of the phase
diagram of the spin-1/2 $J_1-J_2$ Heisenberg model on the honeycomb
lattice have obtained conflicting results. The central controversial
point is the existence and nature of magnetically disordered phases
when the N\'eel order becomes unstable as increasing the frustration
$J_{2}/J_{1}$. There is a growing
consensus\cite{Fouet,Mulder,Clark,Mosadeq,Mezzacapo,Bishop,Albuquerque,Reuther,Li_2012_honeyJ1-J2-J3}
that a magnetically disordered region should appear. However, the
nature of this region is still not clear with different approaches
giving different results. An early ED work by Fouet {\it et
al.}\cite{Fouet} first claimed that a GSL might appear in the region
$J_{2}/J_{1}\approx 0.3-0.35$, and for $J_{2}/J_{1}\approx 0.4$ the
system might be in favor of the staggered dimer VBC. A recent ED
study by Mosadeq {\it et al.}\cite{Mosadeq} has claimed that a
plaquette valence bond crystal (PVBC) might exist in the region
$0.2<J_{2}/J_{1}<0.3$, and a phase transition from PVBC to the
staggered dimer VBC exists at a point of the region $0.35\leq
J_{2}/J_{1}\leq 0.4$. However, a more recent ED work by Albuquerque
{\it et al.}\cite{Albuquerque}, which has treated larger system
sizes, has been unable to discriminate whether this magnetically
disordered region corresponds to PVBC with a small order parameter
or a GSL. It is possible that the PVBC may just come from the finite size
effects.\cite{Mezzacapo} For larger $J_{2}/J_{1}$, it has been also
hard to discriminate the staggered dimer VBC with spiral phases,
since ED is especially difficult to treat the incommensurate
behavior of spin correlations due to the small lattice sizes.

There are two recent studies of this model using VMC with different
variational wave functions. Clark {\it et al.}\cite{Clark} have used
Huse-Elser states and resonating valence bond (RVB) states, and
claimed that a GSL appears in the region $0.08\leq J_{2}/J_{1}\leq
0.3$; a dimerized state which breaks lattice rotational symmetry for
$J_{2}/J_{1}\gtrsim 0.3$. However, a more recent work by Mezzacapo
{\it et al.}\cite{Mezzacapo} using an entangled-plaquette
variational (EPV) ansatz have obtained lower energy estimates, and
claimed that in the magnetically disordered region $0.2\lesssim
J_{2}/J_{1}\lesssim 0.4$, the PVBC order parameter vanishes in the
thermodynamic limit. Therefore, the PVBC may just come from the
finite size effects. Since the $Z_3$ directional symmetry breaking
order parameter has not been considered in this paper, it is still
not clear that the lattice rotational symmetry is broken or not in
the region $0.2\lesssim J_{2}/J_{1}\lesssim 0.4$.

In a recent study using PFFRG\cite{Reuther} the authors have obtained that within
the magnetically disordered region, for larger $J_{2}/J_{1}$, there
is a strong tendency for the staggered dimer ordering; for low
$J_{2}/J_{1}$, both of plaquette and staggered dimer responses are
very weak. A further recent study using CCM\cite{Bishop} has got a
more quantitative magnetically disordered region: $0.207\pm
0.003<J_{2}/J_{1}<0.385\pm 0.010$, in which the PVBC phase has been
reported. However, the ground state within $0.21\lesssim
J_{2}/J_{1}\lesssim 0.24$ is hard to be determined using this
approach.

The other controversial point is the form of the magnetic long range
order when $J_{2}/J_{1}$ exceeds the magnetically disordered region.
There are two proposals: the anti-N\'eel order\cite{Bishop} or the
spiral order. It is difficult to get a conclusion by ED since it is
hard to treat the incommensurate spin correlations due to small
lattice sizes.\cite{Albuquerque} Both of the recent SE\cite{Oitmaa}
and PFFRG\cite{Reuther} studies have not found any evidence for the
existence of the anti-N\'eel order and concluded that the spiral
state should be the stable ground state. However, both of the VMC
with EPV ansatz\cite{Mezzacapo} and the CCM\cite{Bishop} studies
support the opposite proposal. Since we are interested in the exotic
disordered phases in the magnetically disordered region and focus on
$J_{2}/J_{1}\in \left[ 0,0.5\right] $, we can not exclude the
possibility that the anti-N\'eel order state exists for
$J_{2}/J_{1}>0.5$.

Due to the existence of strong quantum fluctuations and frustration,
the spin-1/2 $J_1-J_2$ Heisenberg model on the honeycomb lattice is
a challenging model which needs further investigation especially for
the nature of the intermediate phase. Unbiased numerical simulations
are still needed, such as the density matrix renormalization group
(DMRG) method.\cite{White_1992,White_2007,Stoudenmire} Recently,
DMRG has been applied to spin-1/2 Kagome Heisenberg
model\cite{Yan,Depenbrock} and square $J_1-J_2$ Heisenberg
model\cite{Jiang}, and obtained GSLs as the ground state. Since
quantum fluctuations are expected to be stronger on the honeycomb
lattice than those on the square lattice, it would be very
interesting to apply DMRG to the spin-1/2 $J_1-J_2$ Heisenberg model
on the honeycomb lattice.

\section*{ACKNOWLEDGMENTS}

We are especially grateful to Hirokazu Tsunetsugu for his
suggestions of this project and numerous enlightening discussions
for numerical calculations. We would like to thank Peng Li for
fruitful discussions. Hao Zhang is supported by Japanese Government
Scholarship from MEXT of Japan. C. A. Lamas is partially supported
by CONICET (PIP 1691) and ANPCyT (PICT 1426).



\begin{thebibliography}{99}
\bibitem{error} The errors in the values of the phase boundaries which are implicit
in the number of significant figures quoted, only reflect the error
in the extrapolation of our finite-size results to the thermodynamic
limit. They are not intended to represent the essentially unknown
errors implicit in the mean field approach, e.g., the differences in
the local magnetization and the ground state energy per unit cell
compared with the QMC results in the unfrustrated limit.
\end{thebibliography}
\end{document}